\documentclass[jkps,preprint,fleqn,showpacs,showkeys]{revtex4}
\usepackage{graphicx}
\usepackage{amssymb}
\usepackage{amsmath}
\usepackage{bm}
\begin{document}
\setcounter{page}{0}

\title{Controllable conditional quantum oscillations and synchronization
 in superconducting flux qubits}

\author{Ai Min \surname{Chen}}
\author{Sam Young \surname{Cho}}
\email{sycho@cqu.edu.cn}
\thanks{Fax: +86-23-65111531}
\affiliation{Centre for Modern Physics and Department of Physics,
Chongqing University,  Chongqing 400044, China}

\begin{abstract}
 Conditional quantum oscillations
 are investigated for quantum gate operations in superconducting flux qubits.
 We present an effective Hamiltonian which describes a
 conditional quantum oscillation in two-qubit systems.
 Rabi-type quantum oscillations are discussed in implementing
 conditional quantum oscillations to quantum gate operations.
 Two conditional quantum oscillations depending on the states of
 control qubit can be synchronized
 to perform controlled-gate operations by varying system parameters.
 It is shown that the conditional quantum oscillations with their frequency
 synchronization make it possible to
 operate the controlled-NOT and -$U$ gates with a very accurate gate
 performance rate
 in interacting qubit systems. Further, this scheme can be
 applicable to realize a controlled multi-qubit operation
 in various solid-state qubit systems.
\end{abstract}

\maketitle

\section{Introduction}
 Quantum gates lie at the heart of realization of quantum computing \cite{Nielsen}.
 Two-qubit gates as well as single-qubit gates have been demonstrated
 in various types of quantum systems such as
 cavity QED \cite{cavityQED}, ion traps \cite{iontraps}, NMR
 \cite{NMR}, quantum dots \cite{QD}, and superconducting charge \cite{charge}
 and flux \cite{Plantenberg}
 qubits .
 To perform qubit operations, normally,
 an electromagnetic field is
 applied such as microwave fields, laser pulses, and oscillating
 voltages,
 which can induce quantum oscillations between qubit states.
 Especially, quantum Rabi oscillations can be applicable to achieve
 a controlled-gate operation because it implies a complete transition between
 two quantum states at the half-period time of the oscillation.
 Of particular interests are to manipulate such quantum oscillations
 between qubit states by applying an external field in association with
 controlled-gate operations.

 A gate operation depending on a control qubit state
 can be performed, which is called {\it conditional gate operation},
 where the target states can be flipped for a control-NOT (CNOT) gate operation.
 It has been experimentally demonstrated  for a pair of
 superconducting qubits in Refs. \cite{charge}
 and \cite{Plantenberg}.
 In the experiments, an individual conditional
 operation has been applied to observe
 a CNOT gate operation in the superconducting charge and flux qubits.
 Also, in Ref. \cite{Gagnebin}, a similar scheme has been theoretically
 discussed
 via time evolution of a two qubit system for an applied pulsed-bias duration.
 In our study, a {\it conditional quantum oscillation} rather than a
conditional
 gate operation is introduced by
 considering an effective interacting two-qubit Hamiltonian
 which can be adjusted within some system parameter ranges.
 We investigate how  conditional quantum oscillations can be
 simultaneously manipulated to perform a controlled-gate operation
 in a controllable and accurate manner.

 To clearly discuss an implementation of conditional quantum oscillations
 to quantum gate operations,
 in this paper,
 we restrict ourselves to a rotating wave approximation (RWA)
 in the presence of applied time-dependent fields,
 which allows us to capture an essential physics for
 controlled quantum gate operations based on conditional quantum oscillations.
 At resonant frequencies,
 conditional Rabi and non-Rabi oscillations are shown to characterize
 the time-dependent dynamics of the two qubit system.
 We discuss a frequency matching condition for
 achievable controlled-gate operations.
 By synchronizing the two oscillation frequencies
 on the matching condition, the CNOT gate operation performance and operation time
 are shown to be controllable to obtain a very accurate gate operation.
 It shows that
 conditional quantum oscillations and their frequency synchronization are
 applicable to various quantum gate operations in solid-state multi-qubit systems
 such as Toffoli and Fredkin gates.
 \section{Model}
Any two-state system can play a role of qubits. In solid-state
systems,
 qubit parameters are controllable.
 In terms of the pseudo-spin-1/2
 language, a single qubit in
 the two states of the basis
 $\{\left|\uparrow\right\rangle,\left|\downarrow\right\rangle\}$ can be described by
 the Hamiltonian
  $H_{i}=\left[(\varepsilon_i+V_i(t))\mbox{\boldmath $\sigma$}^z_i
  - \Delta_i \mbox{\boldmath $\sigma$}^x_i\right]/2$,
 where $\mbox{\boldmath $\sigma$}$'s are the Pauli matrices with the identity matrix
 $\mbox{\boldmath $\sigma$}^0$.
 $\varepsilon_i$ and $\Delta_i$ correspond to the energy difference
 and transition amplitude between the two states of the qubit $i$, respectively.
 $V_i(t)$ is responsible for an interplay between the states of the qubit $i$ and
 a time-dependent applied field.
 In the basis
 $\{\left|\uparrow\uparrow\right\rangle,\left|\uparrow\downarrow\right\rangle,
 \left|\downarrow\uparrow\right\rangle,\left|\downarrow\downarrow\right\rangle\}$,
 let us consider
 the Hamiltonian \cite{Majer,Plantenberg}
  of interacting two qubits $(i\in \{A,B\})$ written by
\begin{equation}
 H=H_A\otimes \mbox{\boldmath $\sigma$}\!_B^0+ \mbox{\boldmath $\sigma$}_A^0 \otimes
 H_B+J\mbox{\boldmath $\sigma$}\!_{A}^{z}\otimes\mbox{\boldmath $\sigma$}\!_{B}^{z},
\end{equation}
 where $J$ is the interaction strength characterizing the interaction between
 the two qubits.
 If the transition amplitudes $\Delta_A$ or $\Delta_B$ become negligible by means of
 controlling system parameters, the Hamiltonian
 has a form of block-diagonal matrix.
 Each block of the matrix Hamiltonian  corresponds to a conditional Hamiltonian.
 For instance,
 the negligible $\Delta_A$ leads to,
 for the states $\left|\uparrow\right\rangle$ and $\left|\downarrow\right\rangle$
 of qubit $A$, the effective qubit $B$ Hamiltonians respectively
 written by

 \begin{eqnarray}
 H_{B:\left|\uparrow\right\rangle}\!\!\!
 &=&\!\!\!\frac{1}{2}\!
 \left[(\varepsilon_B\! +\! 2J +\! V_B(t))\mbox{\boldmath $\sigma$}\!_B^z
 \!+\!(\varepsilon_A \!+\! V_A(t))\mbox{\boldmath $\sigma$}\!_B^0
 \!-\! \Delta_B \mbox{\boldmath $\sigma$}_B^x\right],
\label{Conditional:H1}
 \\
 H_{B:\left|\downarrow\right\rangle}\!\!\!
 &=&\!\!\!\frac{1}{2}\!
 \left[ (\varepsilon_B\!  -\! 2J +\! V_B(t))\mbox{\boldmath $\sigma$}\!_B^z
 \!-\!(\varepsilon_A\! +\! V_A(t))\mbox{\boldmath $\sigma$}\!_B^0
 \!-\! \Delta_B \mbox{\boldmath $\sigma$}\!_B^x\right].
\label{Conditional:H2}
 \end{eqnarray}

 Equations (\ref{Conditional:H1}) and (\ref{Conditional:H2}) show that
 the time-dependent dynamics of the system can be understood in association with a
 combination of {\it two conditional quantum oscillations of the qubit $B$}.
 Away from the qubit degeneracy point, where the qubit energy
 $\varepsilon_A$ ( transition amplitude $\Delta_A$)
 becomes relatively bigger (smaller) than other parameters,
 such conditional quantum oscillations can be achievable \cite{Plantenberg}.

 \section{ Conditional quantum oscillations}
 In the absence of time-dependent applied fields $V_i(t)=0$ ($i \in \{A,B\})$,
 the time-independent conditional Hamiltonians generate
 the time evolution of the states of
 qubit $B$ through the Schr\"odinger equation
 $i \partial_t \left| \psi_{B:s}(t) \right\rangle
 =H^{(0)}_{B:s}\left|\psi_{B:s}(t)\right\rangle$.
 For the two conditional time evolutions of the qubit states, at
 time $t$, the states are written by
 $ \left|\psi_{B:s}(t)\right\rangle = G_{B:s}(t) \left|\psi_{B:s}(0)\right\rangle$,
 where $G_{B:s}(t)=U^{-1}(\eta^{(0)}_{B:s}) \exp\left[-i {\tilde
 H}^{(0)}_{B:s}t\right]U(\eta^{(0)}_{B:s})$
 with ${\tilde H}^{(0)}_{B:s}=U(\eta^{(0)}_{B:s})
 H^{(0)}_{B:s} U^{-1}(\eta^{(0)}_{B:s})$.
 Here, $U(\eta)$ is a unitary transformation matrix which makes the
 conditional Hamiltonians $H^{(0)}_{B:s}$ diagonal respectively with
  $\eta^{(0)}_{B:\left|\uparrow\right\rangle/\left|\downarrow\right\rangle}
  = \tan^{-1}\Delta_B/\left(\varepsilon_B \pm 2 J\right)$.
 The unitary transformation matrix is written by $ U(\eta)
  = \left( \begin{array}{cc}
   \cos\frac{\eta}{2} & -\sin\frac{\eta}{2} \\
              \sin\frac{\eta}{2} & \cos\frac{\eta}{2} \end{array}
              \right)$.
 These unitary transformations give the eigenvalues of the conditional
 Hamiltonians, i.e., the eigenvalues of the system,
 without time-dependent external fields:
 $\varepsilon^{(0)}_{\pm:\left|\uparrow\right\rangle}
  =
  \frac{1}{2}\left( \varepsilon_A \pm
  \left[(\varepsilon_B+2 J)^2+\Delta_B^2\right]^{1/2}\right)$
  and
 $ \varepsilon^{(0)}_{\pm:\left|\downarrow\right\rangle}
  =
  -\frac{1}{2}\left( \varepsilon_A \mp
  \left[(\varepsilon_B-2 J)^2+\Delta_B^2\right]^{1/2}\right)$.

 In fact, the two conditional time evolutions of the one qubit can
 characterize the dynamics of two qubit system.
  If the initial state of qubit $B$ is chosen as an arbitrary state
  $\left|\psi_{B:s}(0)\right\rangle
  =a\left|\uparrow\right\rangle+b\left|\downarrow\right\rangle$
  with $a^2+b^2=1$,
  the occupation probabilities of the states
  $\{\left|\uparrow\uparrow\right\rangle,\left|\uparrow\downarrow\right\rangle,
 \left|\downarrow\uparrow\right\rangle,\left|\downarrow\downarrow\right\rangle\}$
  at time $t$
 are obtained as
  $P_{s \uparrow}(t)=1-P_{s \downarrow}(t)
  =a^2+[b^2-(a\sin\eta^{(0)}_{B:s}
  +b\cos\eta^{(0)}_{B:s})^2
  ]\sin^2[\Omega^{(0)}_{B:s} t/2]
  $ for the $\left|s\right\rangle$ states  of qubit $A$,
  where the conditional oscillation frequencies are respectively denoted by
  the qubit Larmor frequencies
  $\Omega^{(0)}_{B:\left|\uparrow\right\rangle/\left|\downarrow\right\rangle}
  =\left[(\varepsilon_B\pm 2
 J)^2+\Delta_B^2\right]^{1/2}$
 corresponding to the resultant energy level spacings.
 (i) For $a=1$ and $b=0$, i.e.,
 $\left|\psi_{B:s}(0)\right\rangle=\left|\uparrow\right\rangle$
 or (ii) for $a=0$ and $b=1$, i.e.,
 $\left|\psi_{B:s}(0)\right\rangle=\left|\downarrow\right\rangle$,
  if $\varepsilon_B=-2J$ $(\varepsilon_B=2J)$ then
 the conditional Rabi oscillation between the states
 $\left|\uparrow\uparrow\right\rangle$
 $(\left|\downarrow\uparrow\right\rangle)$ and
 $\left|\uparrow\downarrow\right\rangle$
 $(\left|\downarrow\downarrow\right\rangle)$ occurs with
 the characteristic frequency $\Omega^{(0)}_{R} = \Delta_B$.
 These conditional Rabi oscillations show that
 an initial state of the qubit $B$ can be in the other flipped state of the
 qubit $B$ at the periodic time $t=2\pi(m-1/2)/\Delta_B$ with
 a positive integer $m$.

 For $\varepsilon_B=2 J$, as an example, the conditional Rabi
 oscillation between the states
 $\left|\downarrow\uparrow\right\rangle$ and
 $\left|\downarrow\downarrow\right\rangle$ occurs with its frequency
 $\Omega^{(0)}_{R} = \Delta_B$,
 while a non-Rabi oscillation  between the states
 $\left|\uparrow\uparrow\right\rangle$
 and
 $\left|\uparrow\downarrow\right\rangle$ takes place
 with the frequency $\Omega^{(0)}_{n\mbox{-}R}
 = \left[16 J^2 +\Delta^2_B\right]^{1/2}$.
 At the time $t=\pi/\Delta_B$, if the states
 $\left|\uparrow\uparrow\right\rangle$
 and
 $\left|\uparrow\downarrow\right\rangle$
 are in their original states, i.e., for instance,
 the non-Rabi frequency becomes
 $ \Omega^{(0)}_{n\mbox{-}R}= 2\Omega^{(0)}_{R}$ $
 (J=\frac{\sqrt{3}}{4}\Delta_B=\frac{1}{2}\varepsilon_B)$ \cite{Gagnebin},
 the two qubit system can perform a CNOT gate operation.
 Then
 one can expect a CNOT gate operation at the half-period times of
 the Rabi oscillation
 because
 the states of target qubit can be
 in their flipped states for the one state of
 control qubit (Rabi oscillation) while they can stay at the original
 states for the other state of
 control qubit (non-Rabi oscillation).
 Therefore,
 adjusting the conditional time evolutions of non-Rabi and Rabi oscillations
 enables to perform a controlled-gate operation.
 We will discuss the details of possible controlled-gate operations
 in the presence of time-dependent applied fields below.

\section{Controlled-gate operations with controllable conditional quantum
 oscillations}

 To implement the conditional quantum oscillations
 to a quantum gate operation,
 suppose that the qubit $A$ is the control qubit and
 the qubit $B$ is the target qubit.
 Let us consider a time-dependent applied field $V_{i}(t)= V_i \cos\omega t$,
 where $V_i$ and $\omega$ are its amplitude and frequency, respectively.
 The time-dependent applied field can give rise to
 an external field-driven conditional Rabi oscillation.
 Recall a unitary transformation $U(\eta^{(0)}_{B})$, where
 $\eta^{(0)}_{B}=\tan^{-1} \Delta_B/\varepsilon_B$.
 Actually, any other unitary transformation can be used for the qubit $B$ such as
 $\eta^{(0)}_{B:s}$.
 The conditional
 Hamiltonians are then transformed as
 ${\tilde H}_{B:s}=U(\eta^{(0)}_{B}) H_{B:s} U^{-1}(\eta^{(0)}_{B})$.
  The basis $\{\left|\uparrow\right\rangle,\left|\downarrow\right\rangle\}$
  of qubit $B$
  are transformed into the basis
  $\{\left|0\right\rangle,\left|1\right\rangle\}$ by
  $\left|\tilde \psi_{B:s}\right\rangle=U(\eta^{(0)}_{B})\left|
  \psi_{B:s}\right\rangle$.

  In order to show clearly a possible
  controlled-gate operation,
  we will employ a rotating wave approximation (RWA) which implies
  that an applied time-dependent field can be a static field
  in a rotating frame.
  The system parameters can also be adjusted to satisfy the regime that
  $\varepsilon_B V_B/\Omega_{B}^{(0)} \ll
  \Omega_{B}^{(0)}$ \cite{Saito},
   $V_A \ll \varepsilon_A$, and $|2 J| \ll V_B$, where
  $\Omega_{B}^{(0)}=\left[\varepsilon_B^2+ \Delta_B^2\right]^{1/2}$.
  Within the approximations, then,
  the conditional Hamiltonians in the rotating frame are obtained
   through $\tilde H_{B:s}^{\rm eff}=-i U(t) \partial_t U^{-1}(t)
   + U(t) \tilde H_{B:s} U^{-1}(t)$
   with $U(t)=\exp\left[i\omega t\, \mbox{\boldmath $\sigma$}\!_B^z/2\right]$ as

 \begin{eqnarray}
 \tilde H_{B:\left|\uparrow\right\rangle}^{\rm eff}
 \!\!\!\!&\simeq&\!\!\!\frac{1}{2}\!
  \left[\left(\Omega^{(0)}_{B}\!+\!\frac{2 J\varepsilon_B}{\Omega^{(0)}_{B}}
  \!-\!\omega\right)\!
  \mbox{\boldmath $\sigma$}\!_B^z + \varepsilon_A \mbox{\boldmath
  $\sigma$}\!_B^0 \!+\!
  \frac{V_B}{2}\!\left(\frac{\Delta_B}{\Omega^{(0)}_{B}}\right)\!
   \mbox{\boldmath $\sigma$}_B^x \right]\!,
\label{RWA:effH1}
 \\
 \tilde H_{B:\left|\downarrow\right\rangle}^{\rm eff}
 \!\!\!\!&\simeq&\!\!\!
\frac{1}{2}\!
  \left[\left(\Omega^{(0)}_{B}\!-\!\frac{2 J\varepsilon_B}{\Omega^{(0)}_{B}}
  \!-\!\omega\right)\!
  \mbox{\boldmath $\sigma$}\!_B^z - \varepsilon_A \mbox{\boldmath
  $\sigma$}\!_B^0
  \!+\!
  \frac{V_B}{2}\!\left(\frac{\Delta_B}{\Omega^{(0)}_{B}}\right)\!
   \mbox{\boldmath $\sigma$}_B^x \right]\!.
\label{RWA:effH2}
 \end{eqnarray}

 To see a quantum gate operation, one can employ a probability amplitude table
 ${\cal U}(t)$ at time $t$.
 For the conditional
 quantum oscillations in two qubit systems, then,
 a two-qubit gate operation can be seen
 in the probability amplitude table ${\cal U}(t)$
 expressed as
\begin{equation}
{\cal U}(t) = 
 \left( \begin{array}{cccc}
  P_{\left|\uparrow 0\right\rangle \leftarrow \left|\uparrow
  0\right\rangle}(t)
     & P_{\left|\uparrow 1\right\rangle \leftarrow
     \left|\uparrow 0\right\rangle}(t) & 0 & 0 \\
      P_{\left|\uparrow 0\right\rangle \leftarrow \left|\uparrow
      1\right\rangle}(t)
     &
     P_{\left|\uparrow 1\right\rangle \leftarrow \left|\uparrow
     1\right\rangle}(t)
     & 0 & 0 \\
   0 & 0& P_{\left|\downarrow 0\right\rangle \leftarrow \left|\downarrow
   0\right\rangle} (t)
     & P_{\left|\downarrow 1\right\rangle \leftarrow \left|\downarrow 0\right\rangle}(t) \\
   0 & 0 &
    P_{\left|\downarrow 0\right\rangle \leftarrow \left|\downarrow
    1\right\rangle}(t)
     & P_{\left|\downarrow 1\right\rangle \leftarrow \left|\downarrow
     1\right\rangle}(t)
    \label{truth:1}
  \end{array}
 \right),
\end{equation}
 where $P_{\beta \leftarrow \alpha}$ denotes the probability that
 an $\alpha$ input state becomes a $\beta$ output state with
  $\alpha$, $\beta$ $\in  \{ \left|\uparrow 0\right\rangle,
 \left|\uparrow 1\right\rangle, \left|\downarrow 0\right\rangle,
 \left|\downarrow 1\right\rangle \}$.
 Within the approximations, by solving  the Schr\"odinger
 equations  of the effective Hamiltonians
 $i \partial_t \left| \tilde \psi_{B:s}(t) \right\rangle
 =\tilde H^{\rm eff}_{B:s}\left| \tilde \psi_{B:s}(t)\right\rangle$
 in Eqs. (\ref{RWA:effH1}) and (\ref{RWA:effH2}),
 we obtain the probabilities as
 \begin{equation}
 P_{\left|s 1\right\rangle \leftarrow \left|s 0\right\rangle}(t)  =
\sin^2 \eta_{B:s}\sin^2\frac{\Omega_{B:s}}{2}t
 \end{equation}
 with the relations
$ P_{\left|s 0\right\rangle \leftarrow \left|s
  0\right\rangle}
 \! =\!
   P_{\left|s 1\right\rangle \leftarrow \left|s
  1\right\rangle}
  \!=\!1\!-\!P_{\left|s 1\right\rangle \leftarrow \left|s
  0\right\rangle}
  \!=\!1\!-\!P_{\left|s 0\right\rangle \leftarrow \left|s
  1\right\rangle},
$
 where  the conditional oscillation frequencies are
$
 \Omega_{B:s}
 \!\! =\!\!\left[\left(
 \omega-\Omega^{(0)}_{B}-2 s J\,  \varepsilon_B/{\Omega^{(0)}_{B}} \right)^2
 \!+\! \left(\Delta_B V_B/ \Omega^{(0)}_{B}\right)^2\!\! /4\right]^{1/2}
$ for $\left|\uparrow\right\rangle/\left|\downarrow\right\rangle =
\pm $. The transformation angles are denoted by $\eta_{B:s}
  =\tan^{-1}\left[\frac{ \Delta_B V_B}{2\,
    \left(\omega-\Omega^{(0)}_{B}-2 s J\,  \varepsilon_B/{\Omega^{(0)}_{B}}
  \right) \Omega^{(0)}_{B}}\right]$.

 There are two resonant frequencies
 (i) $\omega =\Omega^{(0)}_{B}
 +2J\,\varepsilon_B/\Omega^{(0)}_{B}$
 and (ii) $\omega = \Omega^{(0)}_{B}
 -2J\,\varepsilon_B/\Omega^{(0)}_{B}$,
 each of which can induce a conditional Rabi oscillation.
 (i) For
 $\omega =\Omega^{(0)}_{B}
 +2J\,\varepsilon_B/\Omega^{(0)}_{B}$,
$ P_{\left|\uparrow 1\right\rangle \leftarrow \left|\uparrow
  0\right\rangle}(t)
 =   \sin^2\frac{\Omega_{R}}{2}t
$
 undergoes a Rabi oscillation
 while
$
 P_{\left|\downarrow 1\right\rangle \leftarrow \left|\downarrow
  0\right\rangle}(t)
 = \sin^2 \eta_{n\mbox{-}R}
  \sin^2\frac{\Omega_{n\mbox{-}R}}{2}t
$
 can do a non-Rabi oscillation,
 where  the conditional Rabi and non-Rabi oscillation frequencies are respectively
 given by

 \begin{eqnarray}
 \Omega_{R} &=& \frac{V_B}{2}\left(\frac{\Delta_B}{\Omega^{(0)}_{B}}\right),
 \label{Rabi}
 \\
  \Omega_{n\mbox{-}R}&=&
 \left[16 J^2 \left(\frac{\varepsilon_B}{\Omega^{(0)}_{B}}\right)^2
   + \frac{V^2_B}{4}
      \left(\frac{\Delta_B}{\Omega^{(0)}_{B}}\right)^2\right]^{\frac{1}{2}}.
 \label{nonRabi}
 \end{eqnarray}

 The amplitude of the non-Rabi oscillation
 is determined by
 $\eta_{n\mbox{-}R}=\tan^{-1}\left[ \Delta_B V_B/8 J\varepsilon_B \right]$.
 The Rabi oscillation has a longer period than
 the non-Rabi oscillation because the non-Rabi frequency is
 larger than the Rabi oscillation frequency,
 $\Omega_{R} <  \Omega_{n\mbox{-}R}$.
 It should be noted that, in the chosen basis, the Rabi
 frequency does not depend on the interaction strength $J$
 within our approximations while the non-Rabi frequency depends on
 the interaction. This shows that if no interaction exists between
 the qubits, in fact, the conditional quantum oscillations
 are not realizable in the basis.
 If one choose other basis, however, a Rabi frequency can
 be dependent of the interaction strength $J$.
 (ii) For the other resonant frequency
 $\omega = \Omega^{(0)}_{B}
 -2J\,\varepsilon_B/\Omega^{(0)}_{B}$,
  the two conditional quantum oscillations exchange their roles each other, i.e.,
$ P_{\left|\downarrow 1\right\rangle \leftarrow \left|\downarrow
  0\right\rangle}(t)
 =   \sin^2\frac{\Omega_{R}}{2}t
$
 undergoes a Rabi oscillation
 while
$
 P_{\left|\uparrow 1\right\rangle \leftarrow \left|\uparrow
  0\right\rangle}(t)
 = \sin^2 \eta_{n\mbox{-}R}
  \sin^2\frac{\Omega_{n\mbox{-}R}}{2}t
$
 can do a non-Rabi oscillation.

 To see clearly a role of  two conditional quantum
 oscillations for CNOT gate operations,
 let us introduce the fidelity $F$ of the probability
 amplitude table for the truth table of CNOT gate
 as $F (t)= \frac{1}{4}\mathrm{Tr}\left[ {\cal U}(t)
 {\cal U}_\mathrm{CNOT}\right]$.
 The fidelity  $F$ and its error $\delta F(t)=1-F(t)$ give
 the estimations of CNOT gate performance and its reliability.
 In terms of the transition probability amplitudes,
 with ${\cal
 U}_\mathrm{CNOT}=\mathrm{diag([0,1,0,0],[1,0,0,0],[0,0,1,0],[0,0,0,1])}$,
 the fidelity is generally given by
 \begin{equation}
  F = \frac{1}{4}
  \Big( P_{\left|\uparrow 0\right\rangle \leftarrow \left|\uparrow 1\right\rangle}
  +
  P_{\left|\uparrow 1\right\rangle \leftarrow \left|\uparrow 0\right\rangle}
  +
  P_{\left|\downarrow 0\right\rangle \leftarrow \left|\downarrow 0\right\rangle}
  +
  P_{\left|\downarrow 1\right\rangle \leftarrow \left|\downarrow 1\right\rangle}
  \Big).
 \label{F}
 \end{equation}
 For the conditional quantum oscillations, from Eq. (\ref{truth:1}),
 the fidelity becomes
 $F(t) = \frac{1}{2}\Big(
   P_{\left|\uparrow 0\right\rangle \leftarrow \left|\uparrow 1\right\rangle} (t)
  +
   P_{\left|\downarrow 0\right\rangle \leftarrow \left|\downarrow
   0\right\rangle} (t)
  \Big)$ because
   $P_{\left|\uparrow 0\right\rangle \leftarrow \left|\uparrow 1\right\rangle} (t)
   =P_{\left|\uparrow 1\right\rangle \leftarrow \left|\uparrow 0\right\rangle}
   (t)$
   and
   $P_{\left|\downarrow 0\right\rangle \leftarrow \left|\downarrow
   0\right\rangle} (t)
   =P_{\left|\downarrow 1\right\rangle \leftarrow \left|\downarrow
   1\right\rangle} (t)$.
 This shows
 that two conditional quantum oscillations with their
 characteristic frequencies $\Omega_{R}$
 and $\Omega_{n\mbox{-}R}$
 determine
 a CNOT operation performance and its reliability.
 For the resonant frequency $\omega
 =\Omega^{(0)}_{B}+2J\varepsilon_B/\Omega^{(0)}_B$,
 the Rabi oscillation $P_{\left|\uparrow 0\right\rangle \leftarrow \left|\uparrow
   1\right\rangle} (t)$ shows that the initial state is in
 its flipped state  at a periodic time
 $t=(m-1/2)\, \tau_R$
 with a positive integer $m$, where $\tau_R=2\pi/\Omega_{R}$
 is the period of Rabi oscillation.
 The non-Rabi oscillation
 $P_{\left|\downarrow 0\right\rangle \leftarrow \left|\downarrow 0\right\rangle}
 (t)$ shows that
 the initial state is in
 its original state at a periodic time
 $t=m\, \tau_{n\mbox{-}R}$,
 where $\tau_{n\mbox{-}R}=2\pi/\Omega_{n\mbox{-}R}$
 is the period of non-Rabi oscillation.
 Performing a CNOT gate operation $(F=1)$ is, in fact, to synchronize
 the periods $\tau_R$ and $\tau_{n\mbox{-}R}$ of the Rabi and non-Rabi oscillations
 by varying the system parameters because, according to the control qubit
 states, the target state
 should be in a flipped state
 (Rabi oscillation: $P_{\left|\uparrow 0\right\rangle \leftarrow \left|\uparrow
  1\right\rangle}=1$) or
 the original state (non-Rabi oscillation:
 $P_{\left|\downarrow 0\right\rangle \leftarrow \left|\downarrow
  1\right\rangle}=0$) at a certain operation time $t_{OP}$.

\section{Synchronization of two conditional oscillations for controlled
 gate operation}

To synchronize the two conditional quantum oscillations for a CNOT
 gate operation, we discuss a frequency matching condition
 between the quantum oscillations.
 If the operation time of CNOT gate is $t^{(1)}_{OP} =\tau_R/2$
 for the first flipped state in the Rabi oscillation,
 a CNOT gate operation can be performed with
 a positive {\it multiple-integer period} of non-Rabi oscillation
 $ n \tau_{n\mbox{-}R} =\tau_R/2$, where $n$ is
 a positive integer.
 Once the frequencies are matched with the condition,
 the CNOT gate is operated  periodically
 at the periodic operation time $t^{(m)}_{OP}=(m-1/2)\,
 \tau_R$ with the $m$-th flipped state of the Rabi oscillation.
 More generally,
 if a CNOT gate operation is performed
 at the $l$-th flipped states in the Rabi
 oscillation, i.e.,
 the operation time becomes $t_{OP} =(l-1/2) \tau_R$
 with a positive integer $l$,
 then
 the matching frequencies are given by
 the relation $ (n+l-1) \tau_{n\mbox{-}R} = (l-1/2)\tau_R$
 because of $ \tau_{n\mbox{-}R} < \tau_R  $
 ($\Omega_R < \Omega_{n\mbox{-}R}$).
 In other words, a $(n+l-1)$ multiple-integer period of non-Rabi oscillation
 matches with the period $(l-1/2)\tau_R$ of Rabi oscillation
 for the CNOT gate operation.
 For $l=1$,
 these matching frequencies are reduced to the relation $ n \tau_{n\mbox{-}R} =\tau_R/2$
 at the first flipped state.
 Consequently, the matching
 frequencies between the conditional non-Rabi and Rabi oscillations
 for a CNOT gate operation
 are given by the relation
 \begin{equation}
  \Omega_{n\mbox{-}R} (V_B)
  =2 \left(\frac{n + l -1}{2 l -1} \right)
  \Omega_{R} (V_B).
  \label{RabiFrequency}
 \end{equation}
 Both the Rabi and non-Rabi oscillations can be tuned by varying
 the amplitude of applied fields $V_B$ for a CNOT gate operation.
 By using the conditional quantum oscillations, therefore,
 a CNOT gate operation can be achievable by synchronizing
 the periods of conditional Rabi and non-Rabi oscillations.
 Note that, for the other resonant frequency
  $\omega =\Omega^{(0)}_{B}
  -2J\varepsilon_B/\Omega^{(0)}_{B}$,
 the matching frequencies in Eq. (\ref{RabiFrequency}) leads to
 another CNOT gate operation with the ideal truth table
 ${\cal U}_\mathrm{CNOT}=\mathrm{diag([1,0,0,0],[0,1,0,0],[0,0,0,1],[0,0,1,0])}$
 because $P_{\left|\uparrow 1\right\rangle \leftarrow \left|\uparrow
  0\right\rangle}=0$ and
  $P_{\left|\downarrow 1\right\rangle \leftarrow \left|\downarrow
  0\right\rangle}=1$ at the operation time $t_{OP} =(l-1/2) \tau_R$.

 When the CNOT gate operation
 has carried out, from Eq. (\ref{RabiFrequency}),
 the amplitudes of the applied time-dependent field $V_B$ are given by
 \begin{equation}
  V^\mathrm{CNOT}_B (n,l) = \frac{2 l -1}{\sqrt{(2n-1)(2n+4 l-3)}}\left(
              \frac{8J\,\varepsilon_B}{\Delta_B } \right).
 \label{VB}
 \end{equation}
 Then as $V_B$ varies a CNOT gate operation is executed for
 $V_B=V^{\rm CNOT}_B(n,l)$ consecutively.
 This implies that the synchronization of the operation time
 $t_{OP}=(l-1/2) \tau_R$ can be achievable by
 tuning the applied time-dependent field $V_B$.
 Also, it is shown that from Eqs. (\ref{Rabi}), (\ref{nonRabi}), and (\ref{VB})
 the CNOT gate operation can be performed with
 {\it possible Rabi and non-Rabi frequencies}
  given as

  \begin{eqnarray}
 \Omega^\mathrm{CNOT}_{R} (n,l)
  \!\!&=&\!\!
 \frac{2 l -1}{\sqrt{(2n-1)(2n+4 l-3)}}\left(\!
              \frac{4J\,\varepsilon_B}{\sqrt{\varepsilon^2_B+\Delta^2_B} }
              \right),
 \\
 \Omega^\mathrm{CNOT}_{n\mbox{-}R} (n,l)\!\!
  &=&\!\!
 \frac{2 (n + l -1)}{\sqrt{(2n-1)(2n+4 l-3)}}\left(\!
              \frac{4J\,\varepsilon_B}{\sqrt{\varepsilon^2_B+\Delta^2_B} }
              \right).
  \end{eqnarray}

 As a result, synchronizing well conditional
 quantum oscillations
 by varying system parameters makes it possible to achieve
 a CNOT gate operation with a very accurate performance rate, which
 can be applied to various types of qubit systems.
 In addition, the operation time can be controlled
 by means of the matching frequencies.

 A CNOT gate is a special case of the controlled-$U$ gate. If the
 conditional non-Rabi oscillation is suppressed to make the states of target
 qubit staying in their original states during the conditional Rabi oscillation,
 the two qubit system can be a controlled-$U$ gate.
 Actually, if $\eta_{n\mbox{-}R}\ll 1$,
  i.e.,
 $\Delta_B V_B \ll 8 J\,\varepsilon_B$,
 the amplitude of the non-Rabi oscillation becomes negligible
 $\sin^2\eta_{n\mbox{-}R}\approx 0$.
 Without the matching frequencies,
 a controlled-$U$ gate operation can then be obtained.
 Another possible way for a controlled-$U$ gate is also
 to be the matching frequencies.
 From Eq. (\ref{RabiFrequency}),
 the amplitude of the non-Rabi oscillation is given by
 \begin{equation}
 \sin^2\eta^{\rm CNOT}_{n\mbox{-}R}
 = \frac{1}{4}\left(\frac{2l-1}{n+l-1}\right)^2.
 \end{equation}
 As $n$ increases, i.e., the amplitude of time-dependent field decreases
 in Eq. (\ref{VB}),
 the amplitude of the non-Rabi oscillation
 can be significantly suppressed, $\sin^2\eta_{n\mbox{-}R}\approx 0$.
 This results in a realization of a controlled-$U$ gate rather than
 a controlled-CNOT gate.
 In the case of $\omega =
 \Omega^{(0)}_{B} \pm 2 J \varepsilon_B/\Omega^{(0)}_B$, then,
 the probability amplitude tables become a truth table of
 controlled-$U$ gate respectively given as

 \begin{eqnarray}
 {\cal U}^{(+)}_{CU}(t) &\simeq&
 \left(\begin{array}{cccc}
   \cos^2\frac{\Omega_{R}}{2}t
   & \sin^2\frac{\Omega_{R}}{2}t & 0 & 0 \\
   \sin^2\frac{\Omega_{R}}{2}t
         & \cos^2\frac{\Omega_{R}}{2}t
         & 0 & 0 \\
   0 & 0 & 1 & 0 \\
   0 & 0 & 0 & 1
  \end{array} \right),
 \\
 {\cal U}^{(-)}_{CU}(t) &\simeq&
  \left(\begin{array}{cccc}
   1 & 0 & 0 & 0 \\
   0 & 1 & 0 & 0 \\
   0 & 0 & \cos^2\frac{\Omega_{R}}{2}t
   & \sin^2\frac{\Omega_{R}}{2}t \\
   0 & 0 &  \sin^2\frac{\Omega_{R}}{2}t
         & \cos^2\frac{\Omega_{R}}{2}t
  \end{array} \right).
 \end{eqnarray}

 As a consequence, if one of conditional quantum oscillations is
 suppressed by controllable system parameters, the two qubit system
 can provide a controlled-$U$ gate.

\section{ Multi-qubit system and conditional quantum oscillation}

For multi-qubit systems, conditional quantum oscillation can be
 realizable if a similar adjustment is made in the system parameters.
 Once conditional quantum oscillations are achieved in multi-quibt
 systems, one may synchronize their characteristic frequencies to
 perform a controlled multi-qubit gate operation such as Toffoli
 and Fredkin gates which are an extension of two-qubit gates to multi-qubit gates.

\section{Conclusion}
 We investigated conditional quantum oscillations in interacting solid-state
 qubit systems. It was shown that a conditional quantum oscillation can be
 achievable in a way of tuning a range of system parameters.
 Synchronizing conditional quantum oscillations by
 varying applied time-dependent fields as well as system parameters
 enables to perform quantum gate operations such as controlled-NOT
 and -$U$ gate operations with a very accurate performance rate and
 adjustable operation time.
 Controlled multiple-qubit gate operations such as Toffoli and Fredkin gates
 can be implemented with conditional quantum oscillations and their synchronization.

\begin{acknowledgments}
We thank Mun Dae Kim for helpful discussions.
 This work was supported by the NSFC under Grant No.10874252
 and Natural Science Foundation  Project of CQ CSTC.

\end{acknowledgments}

\end{document}